# Terahertz response of patterned epitaxial graphene

C. Sorger[1], S. Preu[1,2], J. Schmidt[3], S. Winnerl[3], Y. V. Bludov[4], N. M. R. Peres[4], M. I. Vasilevskiy[4], H. B. Weber[1,*]

[1] Chair for Applied Physics, Friedrich-Alexander-University of Erlangen-Nürnberg, D-91058 Erlangen, Germany
[2] Department of Electrical Engineering and Information Technology, Technical University Darmstadt, D-64283 Darmstadt, Germany
[3] Institute of Ion Beam Physics and Materials Research, Helmholtz-Zentrum Dresden-Rossendorf, D-01314 Dresden, Germany
[4] Department of Physics and Center of Physics, University of Minho - P-4710-057, Braga, Portugal

We study the interaction between polarized terahertz (THz) radiation and micro-structured large-area graphene in transmission geometry. In order to efficiently couple the radiation into the two-dimensional material, a lateral periodic patterning of a closed graphene sheet by intercalation doping into stripes is chosen, yielding unequal transmittance of the radiation polarized parallel and perpendicular to the stripes. Indeed, a polarization contrast up to 20% is observed. The effect even increases up to 50% when removing graphene stripes in analogy to a wire grid polarizer. The polarization dependence is analyzed in a large frequency range from < 80 GHz to 3 THz, including the plasmon-polariton resonance. The results are in excellent agreement with theoretical calculations based on the electronic energy spectrum of graphene and the electrodynamics of the patterned structure.

Graphene's remarkable electrical and optical properties [1,2] make it attractive for designing optoelectronic devices achieved within an atomic layer thickness. Transparent in a broad optical frequency range, with a loss of only 2.3% of incident radiation graphene can be a good conductor under doping or electrostatic gating. The tunability of its conductivity, Fermi energy and, consequently, plasma oscillations offers a broad range of potential applications in the terahertz (THz) spectral range [1]. In this work we demonstrate that patterned graphene can act as a grating producing highly polarized THz radiation. The polarization effect is based on the coupling of the incident electromagnetic wave to surface plasmon-polaritons (SPPs) supported by doped graphene.

We opt for epitaxial graphene because it can be grown continuously and homogeneously on large areas on a THz transparent highly resistive silicon carbide (SiC) substrate with epitaxial control, both as n-type [2] and p-type layers [3-5]. Lithographic patterning allows for the formation of plasmonic gratings in order to enhance the interaction with light. We focus on patterns with a periodicity that is substantially smaller than the free-space THz wavelength (up to 3 orders of magnitude). Such plasmonic gratings are an important tool to manipulate light-matter interactions [1,6], since they allow for the wave vector matching between the incident electromagnetic wave and SPPs. An additional advantage of the graphene/SiC system is that it allows for growing and even combining in one sample different graphene species, for instance, monolayer and quasi-freestanding bilayer graphene (QFBLG)[7]. In various materials, plasmonic enhancement has yet been utilized over more than three orders of magnitude in frequency, ranging from the near infrared (telecom band, 193 THz) [8,9] down to the THz (100 GHz- 10 THz) range [10,11]. In graphene, recent studies at mid- infrared and THz frequencies have revealed that graphene ribbons and discs exhibit pronounced plasmonic effects [12-14]. The phenomena are particularly rich in the spectral vicinity of the SPP resonance. Simply spoken, below the SPP resonance frequency electrons can follow the electrical field, whereas they remain stationary above this frequency, associated with graphene's high transparency.

As a first approximation to the problem, we note the plasma frequency of a continuous two-dimensional conductor like graphene [15]

$$\omega_{pl}^{(2D)}(k) = \sqrt{\frac{n^{(2D)} k\, e^2}{2\, m^* \epsilon_0 \epsilon_r}} \,, \tag{1}$$

with $m^* = t_\perp/(2 v_F^2)$ denoting the effective mass for parabolic dispersion like in bilayer graphene, and $m^* = \hbar\sqrt{\pi n^{(2D)}}/v_F$ being its analogue for monolayer graphene. In this notation, $n^{(2D)}$, $v_F$ and $t_\perp$ are, respectively, the charge density, the Fermi velocity and the interlayer coupling of bilayer graphene [16]. If graphene's conductivity is periodically modulated (in our case we use a stripe pattern with periodicity $D$), then the diffraction of the normally incident electromagnetic wave on periodicity results in the fact that the reflected and transmitted waves are constituted of an infinite series of spatial harmonics with in-plane vectors $k_l =$

$\frac{2\pi l}{D}$, where $l = (-\infty, \infty)$ is an integer. The SPP resonance frequency $\Omega_l$ (corresponding to $l$-th harmonics) is then estimated as

$$\Omega_l = \frac{4}{D}\sqrt{\frac{\alpha \pi l c d E_F}{\hbar(\varepsilon_{vac}+\varepsilon_{SiC})}} = \omega_{pl}^{(2D)}(k_l)\sqrt{\frac{d}{D}}, \tag{2}$$

where $c$ is the velocity of light, $\alpha$ denotes the fine-structure constant, $\hbar$ is the Planck constant, $\varepsilon_{vac}$ and $\varepsilon_{SiC}$ are the dielectric constants of vacuum/air and the SiC-substrate, respectively. The plasma resonance is broadened by the momentum relaxation rate $\gamma$. We have developed a theory of the transmission near the SPP resonance employing a full electrodynamics simulation [17]. It entirely takes into account the band structure of graphene and its specific characteristics (parametrized by the charge density $n$ and the momentum relaxation rate $\gamma$), as well as the system geometry (period $D$ and stripe width $d$) and the SiC substrate (thickness $H$ and dielectric function $\varepsilon_{SiC}$) [23].

We use commercially available semi-insulating hexagonal SiC(0001) as a substrate for the subsequent growth of epitaxial graphene [2]. For periodic patterning we chose a minimum invasive technique that yields n-type and p-type epitaxial graphene side-by-side within a continuous graphene sheet by local intercalation [5]. The resulting pattern into p- and n-doped stripes is sketched in Fig. 1a. A closer look reveals that the hydrogen intercalation patterning chosen here [3,4] affects also the interface layer (buffer layer [2]) in between monolayer graphene and its substrate SiC. Pristine, n-type monolayer graphene (MLG displayed in blue in Fig. 1a, charge carrier density $n \approx 10^{13}$ cm$^{-2}$ with buffer layer underneath in black) are merged with so-called quasi-freestanding bilayer graphene [3,4] (QFBLG displayed in red in Fig. 1a, where the buffer layer has been converted into a second graphene sheet, $p \approx 10^{13}$ cm$^{-2}$). Although the system forms a closed carbon bilayer sheet and differs only by a small amount of hydrogen, a materials contrast is introduced, as shown in scanning electron micrographs (SEM) in Fig. 1b-d. Previous experiments have shown that the lateral contact between both graphene types behaves ohmic (i.e. linear IV characteristics) due to the absence of a band gap in graphene. It is certainly a nontrivial question whether this subtle modulation is capable of coupling the electrons in this two-dimensional graphene sheet to electromagnetic radiation. Using a scheme that has been developed in reference [5], we patterned large-area graphene into p- and n-type stripes to define plasmonic gratings, with a periodicity $D$ = 5 µm and QFBLG-width $d$ = 2.7 µm for sample PN1.

The dependence of sample transmittance upon frequency in the THz range was measured by using a continuous-wave (CW) as well as a pulsed THz time domain setup (TDS: Time Domain Spectroscopy). A n-i-pn-i-p superlattice photomixer is used as a THz source with tunable frequency for the CW measurements [18]. The polarized THz signal is focused onto the graphene:SiC sample using parabolic mirrors. The sample is mounted in the center of a rotation stage. While rotating the sample, the projection of the stripes onto the THz field is altered, resulting in an angle-dependent transmitted power, $P(\alpha)$, similar to a rotating wire grid polarizer [17]. In particular, p-polarization and s-polarization can be selected, being defined as electric field components parallel or perpendicular to the periodicity of the grid, respectively (see Fig. 1f). From the transmitted power, $P(\alpha)$, we extract the visibility $V$ of p-polarized vs. s-polarized configuration,

$$V = (P_p - P_s)/(P_p + P_s) = (T_p - T_s)/(T_p + T_s), \tag{3}$$

where $T_i = P_i/P_{ref}$ is the transmission coefficient, and $P_{ref}$ is the transmission through the empty setup. The advantage of measuring the visibility is that no reference spectra, $P_{ref}$, are required. A Golay cell detector is used for direct detection. Due to limited dynamic range at higher frequencies, CW data were only taken below 1 THz at ambient conditions.

TDS measurements were carried out in a dry nitrogen purged environment in order to extend the frequency range towards $\approx$ 3 THz. An 800 nm (Ti:Sapphire system, pulse duration 50 fs) driven large area emitter (LAE) [19] was used to generate highly polarized THz radiation. The transmitted THz-pulses were detected in the time domain by the electro-optic sampling method. The frequency spectrum is obtained by Fourier-transformation of the main transmitted pulse only (single pass transmittivity, inset of Fig. 2b, solid box) or including the first reflection (inset of Fig. 2b, dashed box). The latter is required for comparison with CW data in order to take into account Fabry-Perot (FP) features. The transmission setup was similar to the CW setup. All measurements were performed at room temperature.

Figure 2a displays the visibility as a function of the frequency of sample PN1. Obviously, the visibility is oscillatory due to FP like multiple reflections at the front and back side of the sample (wafer thickness H = 490 µm, $\varepsilon_{SiC} \approx 10$ [20]) (see inset). It oscillates between $\approx 0.01$ and $\approx 0.12$ resulting in a polarization contrast $C = (T_p-T_s)/T_p$ up to 20%. The visibility in the maxima decreases with increasing frequency, which indicates the approximation to the SPP resonance.

Hence, for getting closer to the resonance, we enhance the spectral region by employing TDS measurements. Single pass TDS and CW measurements for the same sample are shown in Fig. 2a without rescaling. The two different data sets match excellently. The visibility drops below zero, which indicates that $T_p$ becomes smaller than $T_s$. At higher frequencies a minimum in the visibility indicates the SPP resonance at $f$ = 2.3 ± 0.3 THz (strictly spoken, the resonance coincides with the minimum only in the limit of $\Omega_1 \gg \gamma$). At this point, the THz radiation and the electron plasma are in resonance: $T_p$ reaches its minimum (due to the fact, that SPP are p-polarized waves), whereas $T_s$ is essentially unaffected. When further increasing the frequency, the visibility rises again and reaches zero at $f \approx 3$ THz, as both transmission components become equal. Beyond this frequency, the sample becomes highly transparent. Whereas a perfect polarizer would have visibility equal to unity, this device reaches visibilities of ~12 %, which is remarkably high given the very subtle materials contrast of the closed two-layer carbon sheet.

In order to enhance this effect, and to simplify the setup, we increased the materials contrast drastically by designing periodic graphene (QFBLG: samples P1-P3 and MLG: sample N2, N3) stripe patterns, separated by areas where the graphene was entirely removed by oxygen-plasma etching (cf. Fig 1f). This is even more similar to a wire grid polarizer, but with an atomically thin metal. The experimental data of the visibility for sample P1 are shown in Fig. 2b, where a geometry ($D$ = 6 µm, $d$ = 3.5 µm) similar to PN1 was chosen. The CW results show essentially the same FP like oscillations, which are limited to frequencies below 450 GHz due to experimental imperfections. As expected, the maximum visibility is significantly higher than for sample PN1. The geometric parameters as well as the charge density $p = 8.4 \times 10^{12}$ cm$^{-2}$ (from Hall-effect studies) are known for sample P1. Hence, the momentum relaxation rate is the only parameter that needs to be adapted for comparison with theory. For $\gamma$ = 20 meV, a perfect match in the given spectral region is achieved. This corresponds to a polarization contrast C up to 30%. As the experimental data of CW and TDS look quite different, it is instructive to consider not only the first recorded pulse (single pass), but also the second pulse that results from signals reflected back and forth at the substrate surfaces (inset of Fig. 2b). The result after Fourier transformation is displayed in Fig. 2b. The inclusion of the second pulse re-establishes the FP oscillations that immediately connect to the CW data. Inclusion of even larger windows (i. e. higher order round trip pulses) in the time domain is not useful as experimental artefacts would then be included such as reflections within the detector crystal of the TDS system.

For an investigation of the resonance, we chose larger spatial periodicities of the stripe pattern (samples P2 and P3). Figure 2c displays CW data of the visibility of sample P2 ($D$ = 20 µm, $d$ = 10 µm). The effect is qualitatively similar to P1, also with a comparable amplitude at low frequencies, but a significant decrease of the visibility towards 1 THz. This indicates the proximity to the resonance, which can be seen in the single pass TDS data of this sample, displayed in the same plot without rescaling. The two different data sets match excellently, and furthermore agree well with the theory. Visibility measurements of a third sample, P3, are displayed in Fig. 2d. Due to the further increased length-scale of the stripe pattern ($D$ = 60 µm, $d$ = 30 µm), the SPP resonance is expected at a lower frequency $\Omega \approx 1$ THz. Similar to a mechanical oscillator, a higher momentum relaxation rate (relative to the resonance frequency) leads to a broadening of the resonance and a shift towards higher frequencies. Indeed, sample P3 can be considered as overdamped.

When carrying out the experiment with wire grid-like patterns, but using n-type MLG instead of p-type QFBLG, the data are very similar. The experiments were carried out on the very same samples, which are labelled N2 and N3 before intercalation, and P2 and P3 after intercalation, respectively. Figures 2c and 2d display the comparison of these data in a joint plot. It becomes obvious that the quasi-freestanding p-type material displays significantly larger visibilities. We assign this difference to the higher charge carrier mobility of the quasi-freestanding bilayer graphene at room temperature [4,21]. The ratio of the momentum relaxation rates agrees with the respective ratio obtained from the DC Hall mobilities for single layer graphene [4,22].

It should be emphasized that other quasi-periodic patterns exist on any macroscopic SiC chip due to step edges of the SiC substrate. They may be randomly oriented (approximately on-axis wafer cut, with a flower-like step edge pattern) or quasi-periodic due to intentional wafer miscut. We now quantify their influence. First, we investigate a SiC chip without any graphene. The CW measurement results in no resolvable visibility (cf. Fig. 2a). In a next step, graphene is grown by thermal decomposition. In addition to the substrate steps, additional layer growth along the step edges is observed, which results in a quasi-periodic pattern. Depending on the homogeneity of the graphene layer on samples with aligned step edges the measured visibility varies significantly. The findings range from no resolvable visibility for a homogeneous coverage with graphene (inset in Fig. 2c) up to a visibility of 25% for aligned, 2-3 µm spaced step edges decorated with a multitude of graphene flavours. From SEM analysis we know that the graphene layer for the samples P2/P3 and N2/N3 was grown very homogeneously (little step edge decoration). As this periodicity can be rotated by a defined angle with respect to the lithographically induced pattern, its effect can be singled out by regarding the full angular dependence of the transmission. Further, we characterized samples, which have no quasi-periodic substrate-patterns but rather spatially randomized step-edges overgrown with graphene (on-axis sample). These samples show no detectable visibility. All these experiments confirm that the effect reported above on samples P1-P3 and N2-N3 is due to the lithographically defined pattern and only to a minor extent due to quasi-periodic sample features. The case is more complicated for the PN sample, as the fabrication process for the periodic intercalation stripe pattern creates a periodic array of voids (cf. Fig. 1b). We quantified the effect of the void array on the visibility prior to intercalation, i.e. in a homogeneous n-type graphene sheet. It turns out that the signal induced by the void pattern is negligible.

After having demonstrated that samples P1-P3 and N2 behave like a wire-grid polarizer (with insulating regions in between) and can qualitatively and quantitatively be described by the full wave simulation, we return to the PN sample. It is different, because it provides only one closed conducting sheet composed of two different materials. For the simulation we assumed a stripe-like periodic pattern of alternating n- and p-doped regions and employed the momentum relaxation rates ($\gamma_p$ = 11 meV, $\gamma_n$ = 26 meV) derived from P2 and N2 as input parameters for the calculation. The result is displayed in Fig. 2a as solid oscilating line, together with the experimental data already discussed above. We achieve an excellent match with the CW data. This finding implies that the p-n junction in graphene (having no band gap) has little relevance to the observed phenomena.

To conclude, we have demonstrated several possibilities to control the THz transmission through an epitaxial graphene layer by purposeful, periodic patterning. It is remarkable that graphene, despite it is atomically thin and almost fully transparent in the visible, can reduce the transmission by 50 % when stripes are patterned in analogy to a wire grid polarizer. When employing only a very subtle materials contrast by periodic intercalation of hydrogen, even a graphene sheet that covers the full area can reach a reduction of transmission as large as 20 %. We achieve excellent, qualitative and quantitative agreement with theory based on the electronic energy spectrum of the graphene and electrodynamics of the periodically patterned structure. By appropriate design of the periodic pattern, the transition from a metallic DC behaviour to the optical regime can be studied in the THz regime, giving access to the plasmon-polariton resonance.

Acknowledgments
The authors thank J. Jobst for fruitful discussions. The research was performed in the framework of the Sonderforschungsbereich 953 "Synthetic carbon allotropes", funded by Deutsche Forschungsgemeinschaft. We acknowledge support from the EC under Graphene Flagship (contract no. CNECT-ICT-604391).

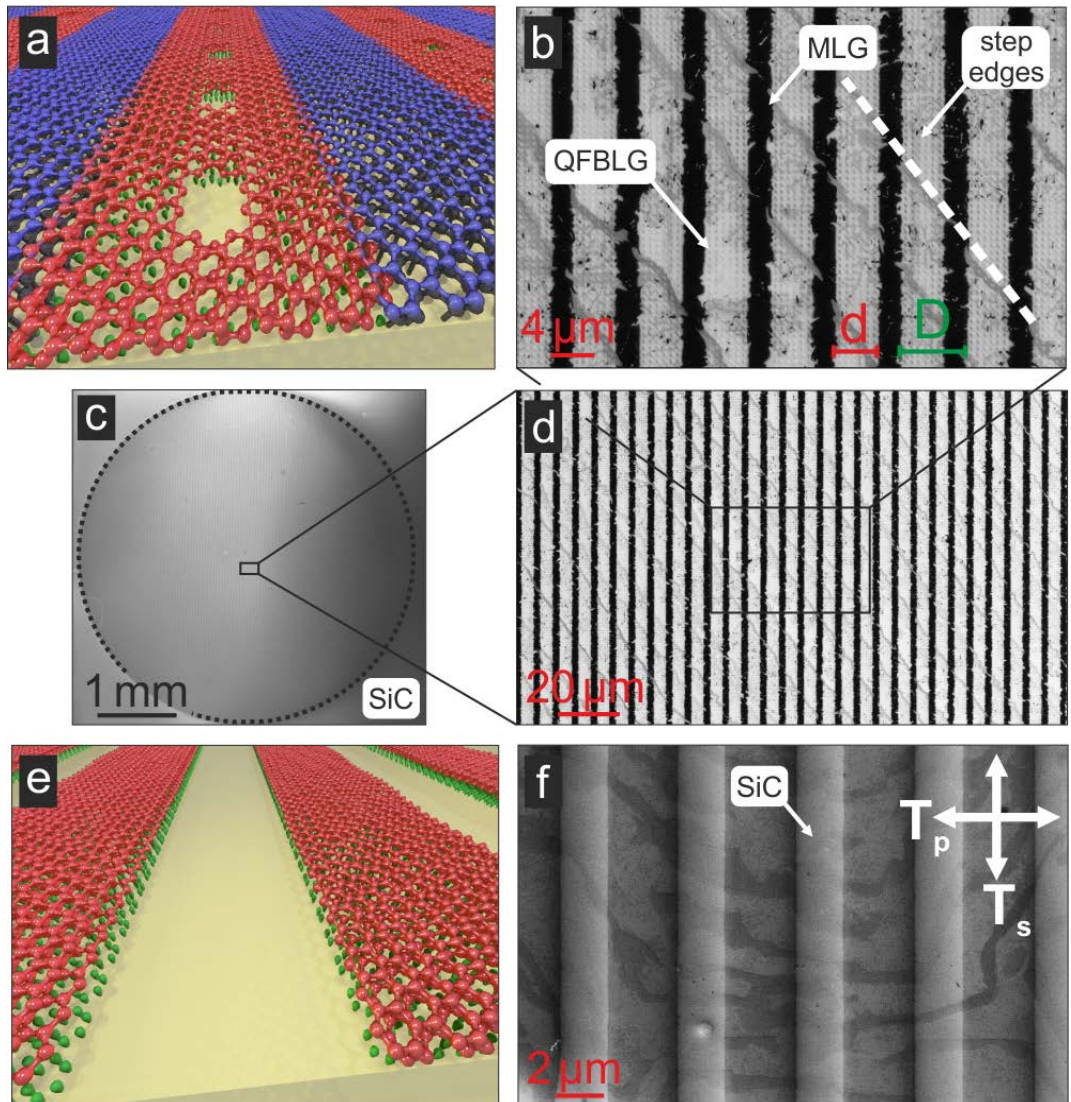

**Fig1.** (a) Artist's view of a strip-pattern of charge-modulation-doping within the epitaxial graphene layer on SiC. The n-type monolayer graphene-sheet (MLG) is drawn in blue, with the buffer layer (black) underneath. Local insertion of hydrogen (green) through predefined voids results in p-type bilayer graphene (QFBLG, red). (b) - (d) Scanning electron micrographs depicting an alternating strip-pattern of QFBLG and MLG (similar to sample PN1). Substrate step edges are oriented in a different angle. (e) Artist's view of an array of isolated strips of QFBLG (samples P1-P3). (f) Scanning electron micrograph of sample P1 depicting isolated strips of QFBLG (light grey). In P1, step edges of the SiC are randomly oriented. Trilayer decoration is visible in SEM, but does not contribute to the experiment.

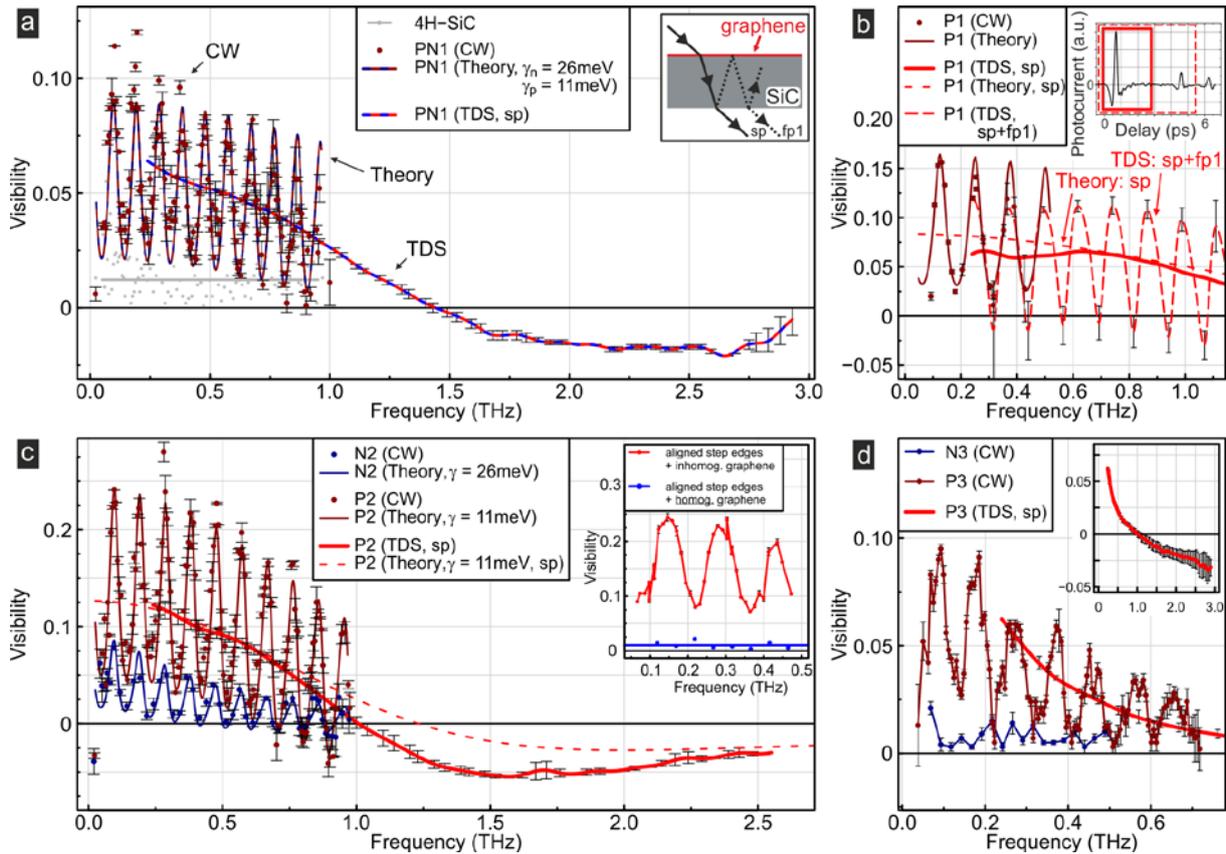

**Fig2.** (a) Experimentally determined visibility of sample PN1 as a function of the excitation frequency as extracted from CW- (dark red symbols) and from TDS-measurements (thick line). Full electrodynamics simulation of the visibility (thinner line) coincide with the experimental CW data. No resolvable signal is recorded from the 4H-SiC substrate prior to graphene growth (grey symbols). In the inset the FP etalon is sketched. (b) Visibility of sample P1 as extracted from the CW- (dark red symbols) and the TDS-setup (solid and long-dashed red lines). Again, coincidence of CW data with theoretical calculations (dark red line) is obtained for a momentum relaxation rate γ = 20 meV. The inset shows the time-domain-signal recorded in TDS-measurements. Fourier transformation of the first detected pulse gives single pass transmission (sp). Second pulse originates from the first round trip inside the sample (fp1). (c) Visibility from CW-data for sample N2 (blue symbols) match the theoretical full wave simulation for single layer graphene (blue line). After intercalation, the resulting sample P2 (CW: dark red symbols; TDS: red line) displays a minimum in the visibility at $f \approx 1.6$ THz. Coincidence with full wave is achieved for γ = 11 meV in the oscillatory region (dark red line). The dashed red line is the simulation in the single pass case. Inset (control experiments): depending on the parameters of graphene growth, high visibilities up to 25% may occur (red symbols), but can be reliably suppressed (blue symbols). (d) Experimental data of the visibility of sample N3 (blue symbols) and P3 (CW: dark red symbols; TDS: red line). The inset shows the TDS data up to $f \approx$ 3 THz. A minimum in the visibility is not observed within the measurement range.